# A Fresnelet-Based Encryption of Medical Images using Arnold Transform

Muhammad Nazeer[1], Bibi Nargis[2], Yasir Mehmood Malik[3], and Dai-Gyoung Kim[1*]

[1]*Division of Applied Mathematics, Hanyang University, Ansan, 426-791, South Korea*
[2]*School of Computer Science, University of Manchester, Manchester, UK*
[3]*Department of Neurology, Rashid Hospital, Dubai, UAE*
[*]dgkim@hanyang.ac.kr

**Abstract:** Medical images are commonly stored in digital media and transmitted via Internet for certain uses. If a medical information image alters, this can lead to a wrong diagnosis which may create a serious health problem. Moreover, medical images in digital form can easily be modified by wiping off or adding small pieces of information intentionally for certain illegal purposes. Hence, the reliability of medical images is an important criterion in a hospital information system. In this paper, Fresnelet transform is employed along with appropriate handling of the Arnold transform and the discrete cosine transform to provide secure distribution of medical images. This method presents a new data hiding system in which steganography and cryptography are used to prevent unauthorized data access. The experimental results exhibit high imperceptibility for embedded images and significant encryption of information images.

## 1. Introduction

Due to recent expansions of information technology, the circulation of medical images among hospitals has become a common trend. Medical images are often distributed for specific purposes such as for teleconferences and interdisciplinary exchanges among medical personnel [1]. The transmission of medical imaging should be carefully treated to ensure reliability and authenticity verification [2]. Privacy is also a critical issue in medical images especially in case of medical jurisprudence where a small manipulation of brain contusion or hairline bony fractures can alter the situation drastically. For interdisciplinary data exchange clarity of image is mandatory such as in multiple sclerosis, a disease of nervous system [3]. If small demyelinating plaques are missed in this disease, whole the diagnostic picture, treatment plan and even prognostic outcome would be altered. Similarly in infectious diseases of chest and abdomen or minimal trauma, if any misinterpretation is found because of imaging defect, clinical outcome would be seriously affected. Thus it is essential to ensure the protection of image information for both legislative and diagnostic reasons. A digital data hiding technique can provide this needed protection by embedding medical information data into other data (called the host or cover data) and has been developed for information security that is strongly based on cryptography and steganography [4]. Steganography and cryptography are both used for data confidentiality. Cryptography is employed for scrambling meaningful information into uncorrelated data keeping the contents of a message secret, whereas steganography hides the significant information under some cover data keeping the existence of a message secret [5]. Each technique enhances the security and privacy for protection of information data. Furthermore, combining both schemes into one system is likely to provide even better security and confidentiality [6]. In other words, steganography prevents an unintended recipient from suspecting the existence of data and the security of the steganography system relies on secrecy of the data encoding system [7]. While cryptography protects messages from unauthorized individual by changing the meaning, steganography techniques enable concealment of the fact that a message is being sent through digital media [8].

One way of data hiding entails the manipulation of the least significant bit (LSB) plane, from direct replacement of the cover LSB's with message bits, to some type of logical or arithmetic combination between the cover image and the amount of information data which needs to be hidden. Several examples of LSB schemes have been presented in [9-11]. Mainly this technique achieves both high capacity and low perceptibility. However, this embedding scheme only overwrites the LSB plane of the cover image with the secret bit stream according to a pseudorandom number generator (PRNG). As a result, some structural asymmetry (never decreasing even pixels and increasing odd pixels when hiding the data) is introduced which makes the disclosure of hidden message very easy even at a low embedding rate using some reported steganalytic algorithms, such as the Chi-squared attack, regular/singular groups (RS) analysis, sample pair analysis [10], and the general framework for structural steganalysis [12].

Most existing steganographic approaches usually assume that the LSB of natural covers is insignificant and random enough, and thus those pixels/pixel pairs for data hiding can be selected freely using a PRNG. However, such an assumption is not always true, especially for images with many smooth regions [10]. Based on the stated researches, we found that natural images usually contain some flat regions as small as 5 x 5, hard to notice. The LSB's in those regions have the values 1 or 0. Therefore, embedding the secret data into these regions will make the LSB of stegano images more and more random, which may lead to visual and statistical differences between cover and stegano images appearing as a noise-like distribution. The pixel-value differencing (PVD)-based scheme [13] is another kind of edge adaptive scheme, which determines the number of embedded bits by the difference between a pixel and its neighbor. This shows the larger the difference is, the larger the number of secret bits can be embedded. Based on our explorative experiments, however, we found that the existing PVD-based approaches cannot make full use of edge information for data hiding, so that the embedded image data does not have a good perceptibility.

Chang et al. extended Iwata et al.'s idea [14-16] and presented a lossless steganographic scheme for hiding secret data in each block of quantized DCT coefficients in JPEG images [17]. This method uses two successive zero coefficients of the medium-frequency components in each block to hide secret data. They further modified the quantization table to maintain the quality of the stegano-image while concealing a higher payload compared with Iwata et al. method. Their scheme achieved reversibility and acceptable image quality of the stegano-image at the same time. However, this scheme can only embed secret bits into the zero coefficients located in the successive zero coefficients in the medium area; non-zero coefficients in the medium area cannot be used. Later, Lin et al. embedded the secret values into the middle frequency of the quantized DCT coefficients, and reduced nonzero values of the quantized DCT coefficients which participate in the data hiding procedure. The aim was to design an adaptive and reversible DCT-based data hiding scheme [18]. Lin and Shiu combined Chang et al.'s [17] method and designed a 2-layer data hiding scheme for DCT-based images. This scheme outperforms Chang et al.'s scheme [16] in hiding capacity but the size of the hidden secret data is still less than 70k bits on average because it retains the reversibility function.

A medical image is distributed among a number of clinicians in telediagnosis and teleconsultation which require maximum information exchange over an unsecure network. Disclosing the information about an important patient's medical condition to general public can be a confidentiality risk. Access to medical information especially image data during transmission which should not be granted to an unauthorized party, demands an important confidentiality measurement. The integrity, on the other hand, demands that images should not be modified in the processes of transmission [19]. However, emerging technologies in field of medical imaging lead to the high demand of protection and confidentiality of the medical information, which must follow strict ethics and legislatives rules. Once the medical images are stolen or abused, the rights of the patients will suffer violation. Tso, H.-K, et al [20] proposes a secret sharing scheme to protect the security of the medical images based on the firstly secret sharing technique proposed by Shamir [21]. According to this technique, no one can obtain any information from one of the shared images unless acquiring the hidden information from all the authorized users. Furthermore, this technique is not efficient for embedding the complete information of secret data in one cover image [22]. Conventionally, information sharing and data hiding are two irrelevant concerns in the field of secure distribution of the information data. In our proposed method, however, a complete information data (e.g., medical image) can be embedded in single cover image (such as patient identity image or face image), rather than distributing in several cover images.

In this study, we proposed a novel technique for data hiding based upon the Fresnelet transform [23] along with applications of discrete cosine transform (DCT) and the Arnold transform [24]. The method based on the Fresnelet transform was initially designed for reconstruction of high resolution digital holography. Irrespective of traditional optical holography, the digital holography has a significant advantage of fast reconstruction process for target object data [25]. Therefore, the Fresnelet transform can be applied to protecting copyrights of digital multimedia contents and data hiding [26]. Besides it is recommended to make use of the Fresnelet transform to complex encryption procedures for providing more security and reliability to information data. Using the Fresnelet transform, an extracted information image data from embedded (cover) image can be obtained with high resolution [27]. And our proposed method in this regard is useful for medical images, as preserving the true resolution of medical imaging system is the key to more accurate understanding of the anatomy, which can support early detection of abnormalities and can increase accuracy in the assessment of size and morphology of organs, effectively [28].

In this paper, we propose a data hiding algorithm to improve security and privacy by integrating both steganography and cryptography with an efficient embedding and extracting process of large size information data. Note that in the proposed method, the Fresnelet transform is employed in order to build a more flexible data hiding system than one using only the Fresnel transform [29] or the wavelet transform [30]. One of the main features of the proposed method is a multi-scale distribution of information on the Fresnelet transform domain yielding robust key parameters for security and privacy. In our presented method, it is not possible to attain the hidden information data without the precise key parameters, even if an attacker perceives the embedding algorithm of the method.

This paper is organized as follows. In Section 2, the Fresnelet transform and Arnold transform are reviewed as the basic transforms in the proposed algorithm. The embedding and extraction schemes based on these basic transforms are presented in Section 3. Numerical simulations and analysis are conducted to evaluate the proposed method in Section 4. In Section 5, conclusions are discussed.

## 2. Basic transforms

*2.1 Fresnelet transform*

A new class of multiresolution bases can be obtained by applying the Fresnel transform to a wavelet basis. The Fresnelet transform has been used for image reconstructions of digital holograms with various parameters related to wavelength, resolution scale, and distance between object and image plane. A review of the Fresnel transform follows from [23].

The Fresnel transform is used to approximately model diffraction phenomena through the propagation of complex waves [29]. The one-dimensional Fresnel transform is defined on a function $f \in L_2(\mathbb{R})$ as the convolution integral

$$\tilde{f}_\tau(x) = (f * k_\tau)(x) \quad \text{with} \quad k_\tau(x) = \frac{1}{\tau} exp\left(i\pi \frac{x^2}{\tau^2}\right) \tag{1}$$

where $\tau > 0$ is the parameter related to wavelength and the distance at a diffracted wave. The two-dimensional Fresnel transform is obtained by convolving with the tensor product of the one-dimensional kernel $k_\tau(x)$. That is, for $f \in L_2(\mathbb{R}^2)$,

$$\tilde{f}_\tau(x,y) = (f * K_\tau)(x,y) \quad \text{with} \quad K_\tau(x,y) = k_\tau(x)k_\tau(y). \tag{2}$$

Notice that the kernel $K_\tau(x,y)$ is separable, so that the useful properties of one-dimensional Fresnel transform are readily extended to two-dimesions. The Fresnel transform has many useful properties. One is the unitary property: processing via the Fresnel transform provides a perfect reconstruction of given data.

The two-dimensional wavelet is obtained from a one-dimensional wavelet by separable extension. The wavelet transform is also defined on $L_2(\mathbb{R})$ as convolution integrals with a two parameter family $\{\psi_{j,l}\}_{j,l \in \mathbb{Z}}$, which forms a Riesz basis for $L_2(\mathbb{R})$, where

$$\{\psi_{j,l}(x) = 2^{j/2}\psi(2^j x - l)\}_{j,l \in \mathbb{Z}}. \tag{3}$$

The simplest example of a wavelet is the Haar wavelet that generates an orthonormal basis for $L_2(\mathbb{R})$. Processing via the wavelet transform provides a perfect reconstruction and a multiresolution decomposition of the data, as well. The properties of wavelet transforms are given in [14]. By applying the Fresnel transform to the wavelet basis, the Fresnelet basis is defined as follows:

$$\{(\psi_{j,l})_\tau^\sim\}_{j,l \in \mathbb{Z}} \quad \text{with} \quad (\psi_{j,l})_\tau^\sim(x) = 2^{j/2}\tilde{\psi}_{2^j\tau}(2^j x - l) \tag{4}$$

with an orthogonal wavelet basis $\{\psi_{j,l}\}_{j,l \in \mathbb{Z}}$, we can have an orthonormal Fresnelet basis. For fixed $\tau$, by letting $\theta_{j,l}(x) = (\psi_{j,l})_\tau^\sim(x)$, we have a Fresnelet decomposition:

$$f = \sum_{j,l} c_{j,l}\, \theta_{j,l} \quad \text{with} \quad c_{j,l} = \langle f, \theta_{j,l} \rangle. \tag{5}$$

Here, the coefficients $c_{j,l}$ are called the Fresnelet coefficients. From the separable nature of the Fresnelet, we may extend the one-dimensional Fresnelet transform to two-dimensional space. In this case, four combinations of tensor products are obtainable:

$$\begin{aligned}
\Theta_{LL} &= (\phi_{j,l})_\tau^\sim(x)(\phi_{j,l})_\tau^\sim(y) \\
\Theta_{LH} &= (\phi_{j,l})_\tau^\sim(x)(\psi_{j,l})_\tau^\sim(y) \\
\Theta_{HL} &= (\psi_{j,l})_\tau^\sim(x)(\phi_{j,l})_\tau^\sim(y) \\
\Theta_{HH} &= (\psi_{j,l})_\tau^\sim(x)(\psi_{j,l})_\tau^\sim(y)
\end{aligned}$$

where $\phi$ and $\psi$ are called the scaling function and wavelet function, respectively, generating low-pass and high-pass filters, respectively. Applying the basis function above to data $f$, we obtain the four types of Fresnelet coefficients:

$$C_{LL} = \langle f, \Theta_{LL}\rangle, \ C_{LH} = \langle f, \Theta_{LH}\rangle, \ C_{HL} = \langle f, \Theta_{HL}\rangle, \ C_{HH} = \langle f, \Theta_{HL}\rangle. \tag{6}$$

The coefficient $C_{LL}$ is the low-passed data and the others are the high-passed data. Fig. 2 shows the Fresnelet coefficients of Eq. 6 applied to the medical information data shown in Fig. 1. In Fig. 2, note that the information image data is scrambled by the transform.

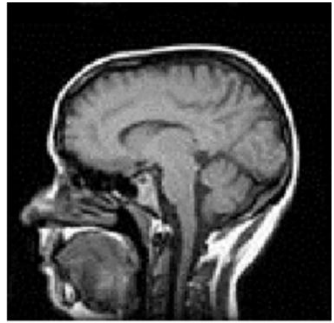

Fig. 1. Medical image for data hiding.

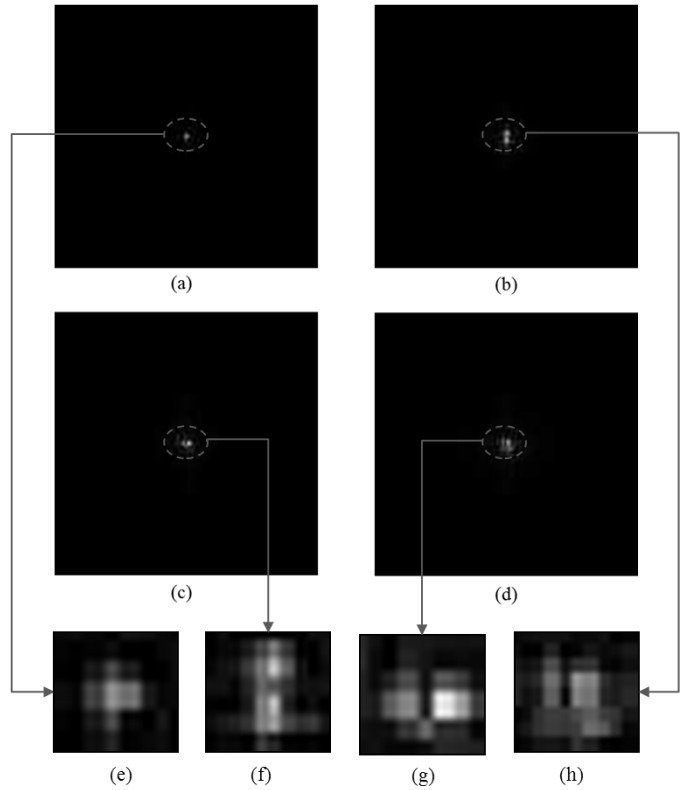

Fig. 2. The magnitude of complex data sets of the Fresnelet coefficients applied to the medical information data shown in Fig. 1: (a) approximate data, (b) horizontal detail data, (c) vertical detail data, (d) diagonal detail data. Moreover, the (e) , (f), (g), and (h) are the zooming parts of marked portion of (a), (b), (c), and (d).

Thanks to the unitary property of the Fresnel transform and wavelet transform, data reconstruction is readily performed by applying the transpose of the forward processing as the inverse Fresnelet transform. Fig. 3 shows the inverse transformed data from Fig. 2, and Fig. 4 shows the result of the reconstruction of the medical information data from the data in Fig. 3. Note here that the proposed transform algorithm depends on the distance and the wavelength key parameters, which are very important for exact reconstruction of the information data. Furthermore, the nature of the reconstruction image is a complex field due to the propagation of complex waves. The complex image is beneficial for medical analysis based on multi-resolution wavelet bases, just as for applications to digital holography [23, 25-27].

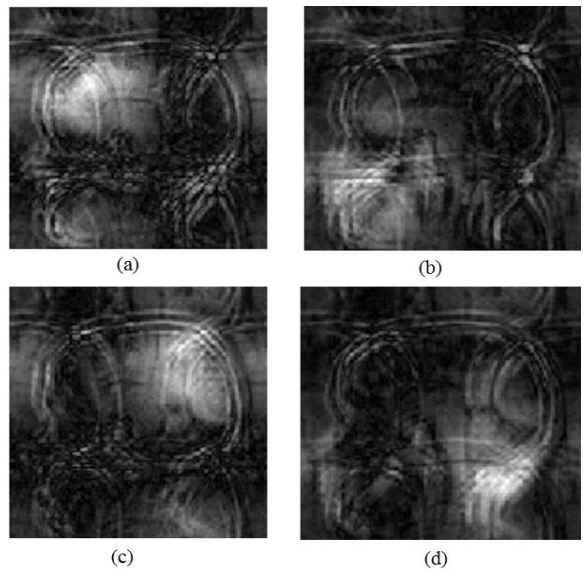

Fig. 3. The magnitude of the inverse Fresnelet transformed data from the four sub-bands of complex data shown in Fig. 2: (a) from approximate data, (b) from horizontal detail data, (c) from vertical detail data, (d) from diagonal detail data.

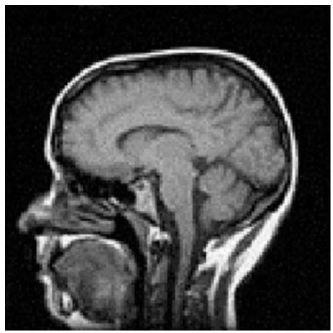

Fig. 4. The reconstruction of the medical information data by combining with the magnitude of four sub-bands of complex data shown in Fig. 3.

For the Fresnelet transform application to the data shown in Fig. 2, the encrypted information data may be covered to protect against a high degree of scrutiny. Also, in the reconstruction phase, the complete information may be acquired by using the exact keys as shown in Fig. 4.

## 2.2. Arnold transform

It is known that the Arnold transform works well in applications for encrypting images [5]. For an $N \times N$ image, the Arnold transform, for example, is given as

$$\begin{bmatrix} x \\ y \end{bmatrix} = \begin{bmatrix} 1 & 1 \\ 1 & 2 \end{bmatrix} \begin{bmatrix} a \\ b \end{bmatrix} \pmod{N} \tag{7}$$

where $(a, b)$ and $(x, y)$ express the pixel co-ordinates of the original and encrypted images, respectively, as shown in Fig. 5.

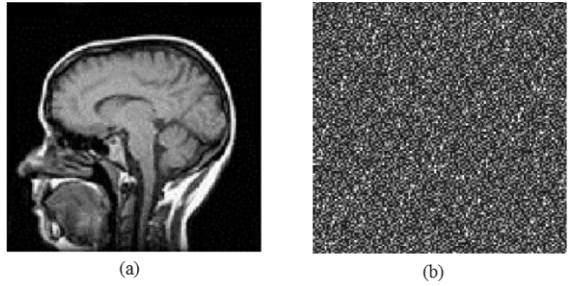

Fig. 5. An example 5 iterations of the Arnold transforms of a medical information image: (a) Medical information image. (b) Information image scrambled by the Arnold transforms.

With the periodic boundary treatment, the image encryption using $n$ iterations of the Arnold transform may be written as

$$I(x, y)^{(k)} = ID(a, b)^{(k-1)} \pmod{N} \tag{8}$$

where $k = 1, 2, \cdots, n$, and $I(x, y)^{(0)} = I(a, b)$. The Arnold transform matrix is given as $D$ in Eq. (7) and $I$ is an $N \times N$ image field. The encrypted image may be inverted by applying $n$ the inverse of the Arnold matrix $D$ $n$ times as follows:

$$I(a, b)^{(k)} = ID^{-1}(x, y)^{(k-1)} \pmod{N} \tag{9}$$

where $(x, y)^{(0)}$ is the pixel of the encrypted image. An original image may reappear after $T$ iterations, depending on the size of the given image. The periodicity $T$ depends on the size of the images, as shown in Table 1.

Table 1. Periodicity of Arnold transforms

| N | 128 | 256 | 480 | 512 |
|---|-----|-----|-----|-----|
| T | 96  | 192 | 240 | 384 |

## 3. Data Hiding Method

*3.1 The embedding process*

The embedding process consists of two phases. First, the Arnold transform is applied on the host image following the wavelet transform. Second, to encrypt the information data, the Fresnelet transform is performed with a single FFT (fast Fourier transform) approach up to the first level by specifying the sampling interval and distance as key parameters. In this case, the sub-bands of information data are of complex data. In the immediate phase, we separate the complex data into real and imaginary parts, respectively. In this way, we get the two sets each of four sub-bands (approximation details, horizontal details, vertical details, and diagonal details). Applying the IDWT (inverse discrete wavelet transforms) on each set separately, we obtain two coded images. These two images of real and imaginary parts are embedded into the same size as the sub-bands of the host image after performing the DCT on each sub-band. Real part which contains important detail of information data is embedded into approximation and horizontal sub-bands of host image, whereas imaginary part is embedded into vertical and diagonal detail of host image. To obtain information embedded data, the IDCT is performed on the each embedded sub-bands of cover image. It is used for distributing the amplitude values of Fresnelet coefficients of information data image of the processed data as shown in Fig. 6.

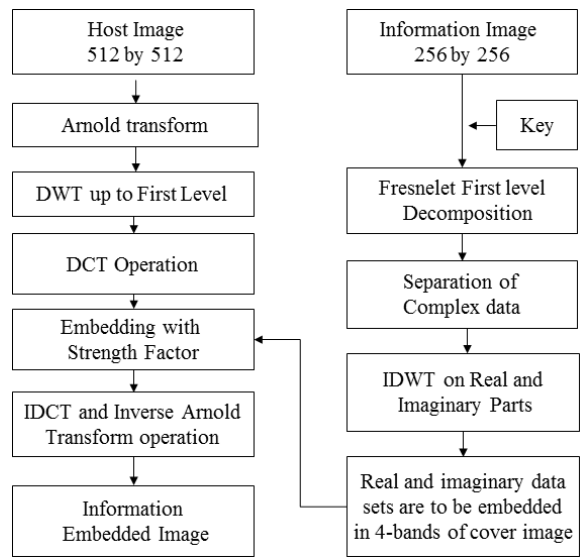

Fig. 6. The schematic diagram of the embedding process.

In this process, the approximate and horizontal detailed parts are modified with real part data, which leads to enhance the strength of the information data, and the vertical and diagonal detailed parts are modified with imaginary part data balance the transparency of the host image. The intensity of the transformed pattern is embedded into the host image with a strength factor. After embedding, the IDWT is performed to write out the image file as an embedded image for digital media handling or internet transmission.

*3.2 The extraction process*

In order to extract the embedded information data, the Arnold transform followed by DWT (discrete wavelet transform) is performed on the embedded data to obtain the sub-bands of the approximate data ($E_a$), horizontal data ($E_h$), vertical data ($E_v$), and diagonal data ($E_d$). The Arnold transform followed by DWT up to the first level is also performed on the original host image, and then each sub-band of the host date is subtracted from the sub-band data $E_a$, $E_h$, $E_v$, and $E_d$, respectively.

Applying DCT on the difference data is performed to extract the information coded data. After DWT is applied, the real and imaginary parts are obtained separately to construct the complex sub-bands for performing the inverse Fresnelet transform. We get the extraction information image in form complex data using the real and imaginary parts which is provided the high resolution for human visual system observation. The whole process of the extraction is shown in Fig. 7.

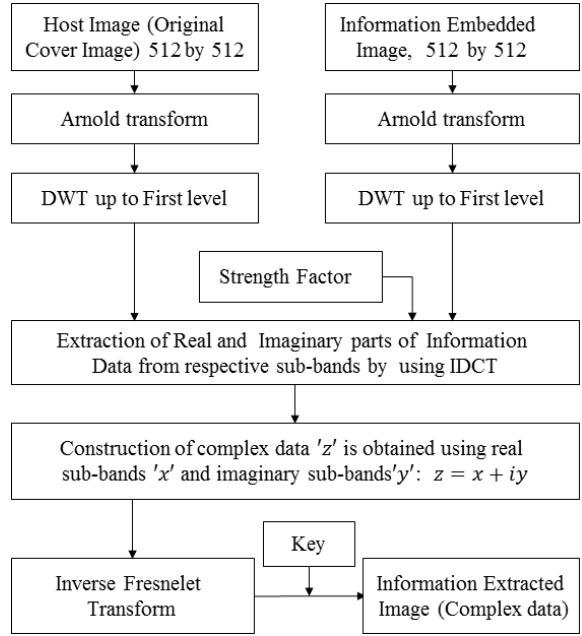

Fig. 7. The schematic diagram of the extraction process.

## 4. Numerical Simulation and Evaluation

*4.1 Simulation and measurement of quality*

In this simulation, the size of all the cover images is $512 \times 512$ and the size of the information images is $256 \times 256$. To analyze the algorithm, MATLAB simulations are performed separately for embedding and extraction of the information data image. The distance $z$ and the wavelength $\lambda$ are key parameters of the Fresnelet transform that improve the security level of the embedded images. In proposed method, we considered a wavelength; $\lambda = 632.8$ nm, sampling interval; $\Delta = 10$ nm, and the distance; $d = 200$ cm. The Arnold transform iteration may also be considered as an additional key parameter. The correlation coefficient (CC) is used to measure the quality of the extracted information data with respect to the original information data before embedding [31]. Specially, ultrasonic detection in medical images is based on amplitude for non-destructive evaluation [32]. In the correlation approach, each signal is correlated with its nearest neighbors for application of matched filtering. In, the A confidence level specifies the relative degree of confidence between two images, with lower to higher values of 0 to 1. The correlation coefficient is given by

$$CC = \frac{\sum_i \sum_j (I_{emb}(i,j)-\bar{I}_{emb})(I_{ext}(i,j)-\bar{I}_{ext})}{\sqrt{\sum_i \sum_j (I_{emb}(i,j)-\bar{I}_{emb})^2}\sqrt{\sum_i \sum_j (I_{ext}(i,j)-\bar{I}_{ext})^2}} \quad (10)$$

where $\bar{I}_{emb}$ and $\bar{I}_{ext}$ are the averages of embedded and extracted information data images, respectively. In addition, the *PSNR* (peak signal noise ratio) is a common estimate of the quality of the embedded image data [33]. Higher values of *PSNR* indicate better imperceptibility of the embedded information data with respect to the cover data:

$$PSNR = 10 \log\left(\frac{255^2}{MSE}\right) \quad (11)$$

$$MSE = \frac{1}{m \times n}\sum_{i=0}^{m-1}\sum_{j=0}^{n-1}[I_{Oi}(i,j) - I_{Ei}(i,j)]^2 \quad (12)$$

where *MSE* is the mean square error between original image $I_{Oi}$ and embedded image $I_{Ei}$ where $I_{Ei}$ is considered a noisy approximation of the $I_{Oi}$ and is shown in Fig. 9-c and Fig. 9-f. Moreover, in order to evaluate the perception quality of human visual system the structural similarity index measure (SSIM), is used to measure the distortion difference two images. The SSIM metrics is designed by modeling any image distortion as a combination of three factors that are loss of correlation, luminance distortion and contrast distortion. The SSIM comparison for measuring the perceptual quality between two images is defined as

$$SSIM(a,b) = l(a,b)c(a,b)s(a,b) \quad (13)$$

$$l(a,b) = \frac{2\mu_a\mu_b + C_1}{\mu_a^2 + \mu_b^2 + C_1} \quad (14)$$

$$c(a,b) = \frac{2\sigma_a\sigma_b + C_2}{\sigma_a^2 + \sigma_b^2 + C_2} \quad (15)$$

$$s(a,b) = \frac{2\sigma_{ab} + C_3}{\sigma_{ab} + C_3} \quad (16)$$

where the $l(a,b)$ is the function of lunminnace comparison to measure the images closeness on the base of mean luminance $\mu_a$ and $\mu_b$ of 2D image. Maximum value of $l(a,b)$ is equal to 1, if and only if $\mu_a = \mu_b$. The second term c$(a,b)$ is used in Eq. (13) to measure the contrast on the base of standard deviation $\sigma_a$ and $\sigma_b$. The maximal value of contrast term as given in Eq. (15) is achieved at $\sigma_a = \sigma_b$. Third term in Eq. (13) is the $s(a,b)$ which measure the structure comparison between the two images. Where, the $\sigma_{ab}$ is the covariance which is useful to analyze the correlation between two images. In the above Eq. (14)-(16) the positive constants $C_1$, $C_2$ and $C_3$ are used, respectively, to eliminate the null denominator. The quality value of SSIM is varied in the positive index of [0, 1]. A value of 1 show that two images are having the same quality and value of 0 shows no correlation between the two images.

*4.2 Evaluation criteria*

In this algorithm, the information image data is embedded in a cover image to increase the security level and improve the imperceptibility of the cover media. For evaluation of the proposed method, the CC factor and SSIM of the information image, whereas the PSNR and SSIM of cover image before embedding and after extraction are analyzed, respectively. The resulting approximately good numeric values show the superior performance of the proposed algorithm. A slight effect is present, mainly due to auto focus nature achieved using Fresnelet sparsity criterion for extraction of information data in form of complex values, as shown in Fig. 8 through 11. The corresponding *PSNR* and *SSIM* are shown in Tables 2 for signature image 1 (the brain image in Fig. 8-a), signature image 2 (the backbone image in Fig. 8-b), and signature image 3 (the USAF target image in Fig. 8-c), respectively. The signature images are high resolution information image data.

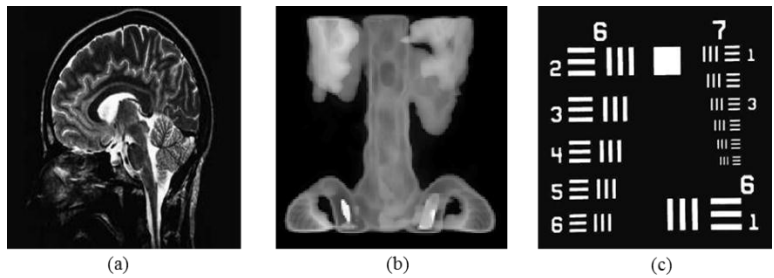

Fig. 8. High resolution signature image data with size: 256 by 256: (a) brain medical image, (b) backbone medical image, and (c) USAF test image.

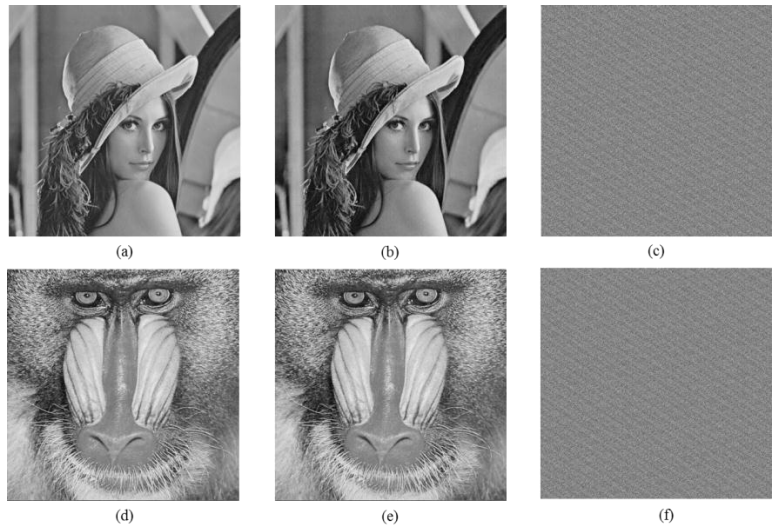

Fig. 9. The cover images embedded with signature image 1 as shown in Fig. 8-a: (a)the original cover image Lena, (b) the cover image Lena embedded with signature image 1 with MSE is 4.3273., (c) the difference between the original cover image Lena and Lena embedded image with signature image 1. (d) the original cover image Mandrill, (e) the cover image Mandrill embedded with signature image 1, (f) the difference between the original cover image Mandrill and Mandrill embedded image with signature image 1 with MSE is 4.3268.

Table 2. PSNR and SSIM of the embedded images: Lena and Mandrill

| Test data | Extraction from embedded image Lena | | Extraction from embedded image Mandrill | |
|---|---|---|---|---|
| | PSNR | SSIM | PSNR | SSIM |
| Signature image 1 | 38.8483 | 0.9982 | 40.8311 | 0.9985 |
| Signature image 2 | 36.0625 | 0.9966 | 36.0638 | 0.9956 |
| Signature image 2 | 36.7419 | 0.9971 | 38.7388 | 0.9976 |

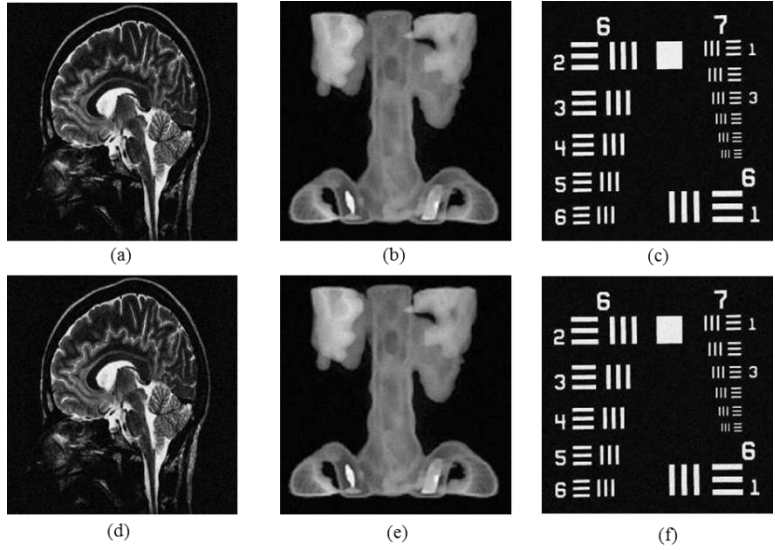

Fig. 11. Extraction of signature images from embedded images of Lena and Mandrill: (a) the signature image 1 extracted from embedded image Lena, (b) the signature image 2 extracted from embedded image Lena, (c) the signature image 3 extracted from embedded image Lena, (d) the signature image 1 extracted from embedded image Mandrill, (e) the signature image 2 extracted from embedded image Mandrill, (f) the signature image 3 extracted from embedded image Mandrill.

Table 3. CC factor and SSIM of signature data extracted from embedded images Lena and Mandrill

| Test Images | Extraction from embedded image Lena | | Extraction from embedded image Mandrill | |
|---|---|---|---|---|
| | CC factor | SSIM | CC factor | SSIM |
| Signature image 1 | 0.9965 | 1.0000 | 0.9947 | 1.0000 |
| Signature image 2 | 0.9978 | 1.0000 | 0.9977 | 1.0000 |
| Signature image 2 | 0.9981 | 1.0000 | 0.9972 | 1.0000 |

The SSIM and CC factor evaluation of Table. 4 show the competency of proposed method. However, it is not appropriate to compare the extraction results of information data with other existing methods due to complex values extraction results of proposed method (usually, existing methods do not extract the secret image in form of complex values).

To further prove our scheme's performance on hiding capacity, we compared our proposed scheme with other existing schemes: to compare the quality of the embedded images and view the degradation caused by embedding a hidden data it is necessary to look at images with a special payload size [34]. Usually, the resulting perceived quality or peak signal-to-noise ratio (PSNR) and payload or bits per pixel (bpp) are the two most commonly-used criteria for evaluating the performance of reversible data hiding techniques [35].

**Table 4. The comparison of image quality using PSNR (dB) and embedding capacity (bits)**

| Cover Images | Lena (512 x 512) | | Baboon (512 x 512) | |
|---|---|---|---|---|
| Measurements | Capacity (bits) | PSNR (dB) | Capacity (bits) | PSNR (dB) |
| Celik et al.'s [18] | 74600 | 38.00 | 15176 | 38.00 |
| Xuan et al.'s [19] | 85507 | 36.60 | 14916 | 32.80 |
| Tian et al.'s [20] | 233067 | 29.97 | 95852 | 29.41 |
| Alattar et al.'s [21] | 173655 | 36.60 | 86264 | 36.60 |
| Wu-Tsai's al.'s [26] | 51219 | 38.94 | 57146 | 33.43 |
| Kamstra et al.'s [22] | 135547 | 35.20 | 103653 | 30.12 |
| Chin-C. C et al.'s [32] | 36850 | 30.34 | 35402 | 26.46 |
| Luo, W et al.'s [27] | 66064 | 38.80 | 68007 | 33.33 |
| Mandal et al.'s [24] | 216000 | **40.92** | 216000 | **40.97** |
| Proposed Method | **263222** | 38.65 | **263222** | 38.65 |

Table. 4 shows the comparison of our results with Celik et al.'s [36] technique by compressing the quantization residuals of pixels instead of compressing LSB planes to obtain more extra space for embedding secret data. The compressed residuals and secret data are concatenated and embedded into the cover image. Xuan et al.'s [37] proposed a scheme based on integer wavelet transform (IWT) and created more space in high frequency sub-bands. Tian al.'s [38] proposed a high capacity reversible data embedding technique that is called difference-expansion (DE) embedding. Alattar al.'s [39] extended Tian's scheme by introducing the differences expansion of a vector to obtain more extra space for embedding secret data. Afterwards, many techniques, such as Kamstra et al.'s algorithm [40], Chin-C, C et al.'s [41], Luo, W., et al.'s [42], and Mandal et al.'s [43] have been propose to increase the embedding capacity and kept the distortion low. But all such methods do not maintained the high resolution images (e.g., medical images etc.) in their extraction phase. However, it is found that hiding capacity of Mandal et al.'s method is higher than other such prevailing techniques, but its hiding capacity is less than our proposed method with cover image size 512 by 512 and payload size 256 by 256, respectively. Moreover, overall performance of our proposed method shows significant performance with other existing schemes under typical circumstances as demonstrated in Table. 4. Besides this, our proposed method is highly capable to preserve the ideal resolution of information data images at their extraction stage. Additionally, multi keys establishment for extraction phase provides the high end confidentiality to concealed information data, whereas appropriate encryption develop the great imperceptibility for embedded image. Note the experimental results presented in Table 4 for Comparison of image quality using PSNR (dB) and embedding capacity (bits) are referred to their original experimental results in their paper. It is the size of payload data (bits) in a cover image that can be modified without deteriorating the integrity of the cover image [35]. However, it is not possible to simultaneously maximize robustness, imperceptiveness, and capacity. Therefore, the acceptable balance of these items must be dictated by the application. For example, an information-hiding scheme may forgo robustness in favor of capacity and low perceptibility, whereas a watermarking scheme, which may not require large capacity or low perceptibility, would certainly support increased robustness [6].

## 5. Conclusion

The present paper proposes a novel data hiding scheme to increase the security and privacy of medical information image data by integrating both steganography and cryptography, based on the Fresnelet transform. The main feature of the proposed method is a multi-scale distribution of information on the Fresnelet transform domain with robust key parameters for security, so that the very small distortion in the cover image provides sufficient privacy for the embedded information. In particular, the Fresnelet decomposition provides robustness to leakage of the energy of information data during the embedding process. For greater security, a combination of the Arnold transform and DCT has been employed in the embedding process. Evaluation of the proposed method has been performed by using histogram analysis. Numerical simulation results have been analyzed with PSNR, SSIM, MSE, and CC factors. Higher values of PSNR, SSIM, and CC factors and smaller value of MSE show the effectiveness of the proposed novel algorithm to achieve great imperceptibility of embedded information image data with required security.

**Acknowledgments**

This work was supported by the National Research Foundation of Korea (NRF) grant funded by the Korea government (MEST) (NRF-2011-0026245).